# Growth dynamics and thickness-dependent electronic structure of topological insulator Bi$_2$Te$_3$ thin films on Si

Yao-Yi Li[1], Guang Wang[1], Xie-Gang Zhu[1], Min-Hao Liu[1], Cun Ye[1], Xi Chen[1], Ya-Yu Wang[1], Ke He[2], Li-Li Wang[2], Xu-Cun Ma[2], Hai-Jun Zhang[2], Xi Dai[2], Zhong Fang[2], Xin-Cheng Xie[2], Ying Liu[3], Xiao-Liang Qi[4], Jin-Feng Jia[1]*, Shou-Cheng Zhang[4], and Qi-Kun Xue[1,2]

[1] *Key Lab for Atomic, Molecular and Nanoscience, Department of Physics, Tsinghua University, Beijing 100084, P. R. China*

[2] *Institute of Physics, The Chinese Academy of Sciences, Beijing 100190, P. R. China*

[3] *Department of Physics, The Pennsylvania State University, Pennsylvania 16802, USA*

[4] *Department of Physics, Stanford University, Stanford, California 94305-4045, USA*

Abstract

We use real-time reflection high energy electron diffraction intensity oscillation to establish the Te-rich growth dynamics of topological insulator thin films of Bi$_2$Te$_3$ on Si(111) substrate by molecular beam epitaxy. *In situ* angle resolved photoemission spectroscopy (ARPES), scanning tunneling microscopy and *ex situ* transport measurements reveal that the as-grown Bi$_2$Te$_3$ films without any doping are an intrinsic topological insulator with its Fermi level intersecting only the metallic surface states. Experimentally, we find that the single-Dirac-cone surface state develops at a thickness of two quintuple layers (2 QL). Theoretically, we show that the interaction between the surface states from both sides of the film, which is determined by the penetration depth of the topological surface state wavefunctions, sets this lower thickness limit.

Traditionally, $Bi_2Te_3$ is known for having the highest figure-of-merit coefficient ZT≈1 among bulk thermoelectric materials [1-3]. Recent theory and experiment reveal that stoichiometric $Bi_2Te_3$ is also a topological insulator (TI) with surface states that reside in its bulk insulating gap that are protected by time-reversal-symmetry[4-6]. The metallic surface states consist of a single Dirac cone at the Γ point and are predicted to exhibit a number of striking electromagnetic properties [7-10], which have recently attracted great attention[4-20]. From the point view of experiment, a major obstacle in the rapidly developing field of TIs is the extreme difficulty in growing intrinsic TI materials, where we define intrinsic by reference to intrinsic semiconductors: its Fermi level ($E_F$) lies in between the bulk conduction band minimum (CBM) and the bulk valence band maximum (VBM) and only intersects the metallic surface state. The reported topological insulators such as $Bi_2Te_3$ and $Bi_2Se_3$ all suffer from a great amount of unwanted bulk carriers [5, 6, 11, 16-19], and an intense bulk electron pocket appears at the Fermi level [5, 6, 11]. With the bulk electronic states at $E_F$, it is difficult to characterize the pristine topological transport property and to use them to develop topological devices that rely only on the behaviors of Dirac fermions.

The $Bi_2Te_3$ crystals studied in recent experiments were grown by melting stoichometric mixture of Bi and Te in a crucible [5, 6, 11]. Local composition fluctuation, which results in a high background carrier density [5, 6], cannot be avoided with this technique. To compensate the background carriers and remove the bulk states from the Fermi level, the materials had to be heavily doped by 0.67% Sn [5]. The situation makes systematic doping control and device gating difficult. Thin films have several advantages: band engineering can be achieved by bipolar or gradient doping, and tunneling junctions as well as heterostructures and superlattices can be fabricated. Growth of $Bi_2Te_3$ and $Bi_2Se_3$ films by molecular beam epitaxy (MBE) and other techniques has been investigated previously [21-23]. With these methods, nominally stoichiometric single crystalline films could easily be prepared. However, discussion of their topological properties was not possible in those studies since the band structure of the films is unknown.

The main purpose of this work is to establish the MBE growth conditions by which intrinsic topological insulator thin films of $Bi_2Te_3$ can be readily obtained. We do this by a systematic study of growth dynamics under various $Te_2$/Bi flux ratios and Si substrate temperatures with reflection high-energy electron diffraction (RHEED). We identify unique Te-rich growth dynamics for preparing intrinsic topological insulator films by the characteristic RHEED intensity oscillations of layer-by-layer MBE growth. *In situ* angle-resolved photoemission spectroscopy (ARPES) and scanning tunneling microscopy (STM) measurements reveal that the as-grown $Bi_2Te_3$ films all exhibit the characteristic intrinsic topological feature: $E_F$ lies in between the bulk conduction band minimum (CBM) and the bulk valence band maximum (VBM) and thus only crosses the metallic surface state. By thickness-dependent band structure measurement, we show that the single-Dirac-cone surface state forms at a thickness of 2 quintuple layers (2 QL), which agrees well with our first principle calculations.

Our experiments were performed in an ultra-high vacuum system that houses an MBE growth chamber, a low temperature scanning tunneling microscope (Omicron) and an angle resolved photoemission spectrometer (GAMMADATA SCIENTA). The base pressure of the system is better than $5\times10^{-11}$ Torr. Si(111)-7x7 substrates were cleaned by standard multi-cycle flashing process [24]. RHEED patterns were used to calibrate the in-plane lattice constant of the $Bi_2Te_3$ films with respect to the Si(111)-7x7 surface, while the RHEED intensity of the (0, 0) diffraction recorded by a CCD camera was used to measure the growth dynamics. The STM images were acquired at 77K using chemically etched polycrystalline W tips. The ARPES spectra were collected with a R4000 analyzer and a VUV 5000 UV source with a monochrometer. The He I (21.218 eV) resonant line was used to excite photoelectrons. The energy resolution and angular resolution of the analyzer were set at 10 meV and 0.2 degree, respectively. Transport measurements were conducted *ex situ* using a standard four-probe, low frequency lock-in method. The MBE-grown thin films were covered with a 10nm-thick tellurium capping layer in UHV before being transferred to atmosphere.

To establish the optimal MBE growth conditions, we start with an analysis of the rhombohedral crystal structure of $Bi_2Te_3$. As schematically shown in Fig. 1a, along the [111] crystallographic direction, the unit cell contains five atomic layers with a stacking sequence of Te(1)-Bi-Te(2)-Bi-Te(1), forming a unique QL. The lattice constant along this direction is 10.17Å, while that on the *a-b* (111) plane is 4.38Å. The bulk and the projected (111) Brillouin zones are shown in Fig. 1b. The QL is terminated by a Te(1) atomic layer in both sides. The interaction between two adjacent QLs is of the van der Waals type, which is much weaker than that between two atomic layers within a QL [4]. As a result, the cleaved surfaces from a bulk crystal are always Te-terminated and have an unreconstructed (1x1)-Te structure [4-5, 18-20]. Inspired by the well-established layer-by-layer growth of GaAs with $As_4$ molecular beam, where the growth unit is always a Ga-As bilayer along [001] direction [25], we anticipate that an ideal MBE growth of $Bi_2Te_3$ with $Te_2$ molecular beam should proceed under Te-rich conditions and in unit of a QL along the [111] direction.

The optimization of our MBE growth of $Bi_2Te_3$ involves a set of growth experiments under different $Te_2$/Bi flux ratios (θ) and substrate temperatures ($T_{Si}$). It was found that the intrinsic insulator could readily be achieved under Te-rich condition (θ=8−20) and when the criterion of $T_{Bi} > T_{Si} \geq T_{Te}$ is satisfied. Here, $T_{Bi}$ and $T_{Te}$ are the temperatures of Bi and Te Knudsen cells, which were used to precisely control the deposition flux (thus the ratio) of Bi and Te. The resulting growth rate is typically ~1/3 QL per minute. In Fig. 1c, we show the RHEED patterns of the initial Si(111)-7x7 surface (upper panel) and the $Bi_2Te_3$ film with a thickness of 4 QL (lower panel). The electron beam incidence is along [11$\bar{2}$] ($\bar{\Gamma} - \bar{M}$) direction. The sharp streaky pattern as seen in the lower panel indicates that the film has an atomically flat surface morphology. A typical large-scale STM image of a 80-nm-thick film grown under the same condition is shown in Fig. 1d. The atomically flat morphology is immediately evident. The 1x1 RHEED pattern, as well as the high resolution STM image with a well-defined 1x1 symmetry (see the insert of Fig. 1d), reveals that the growing front surface during and after deposition is the Te-terminated Bi$_2$Te$_3$(111)-(1x1) surface [18-20], the same as that obtained by cleaving bulk crystals [5, 6]. The steps shown in Fig. 1d are all 10.17 Å in height, corresponding to 1 QL. As expected, the growth on Si(111) proceeds along the [111] crystallographic direction.

Figure 1e shows the intensity evolution of the (0, 0) diffraction (the central streak) versus growth time under various Te/Bi flux ratios and substrate temperatures ($T_{Si}$). In all nine sets of growth parameters depicted in Fig. 1e, there is a persistent RHEED intensity oscillation, which is characteristic of layer-by-layer growth. One period of oscillation corresponds to deposition of a quintuple layer of Bi$_2$Te$_3$. In this sense, the growth is better described as QL-by-QL growth. To further demonstrate this, in some runs of our experiment we intentionally interrupted the growth when 1/4 QL had been deposited. The STM images from as-deposited surface show the formation of 2D islands on the flat terraces, and all with a height of 1 QL. This confirms that the growth unit is 1 QL.

The second important observation in growth dynamics is that the growth rate is only dependent of the Bi flux under our growth conditions. As shown in the upper panel of Fig. 1e, the growth rate increases by a factor of 2.04 when the Bi cell temperature is increased from 763 K to 783 K. This temperature change corresponds to a change of the Bi beam equivalent pressure by a similar factor of 2.08. On the other hand, there is basically no change in growth rate even with a change of the Te beam equivalent pressure by a factor of 2.65, when $T_{Te}$ is increased from 523 K to 543K. We also found that if only the Te cell shutter is open so that only the Te beam is exposed to the substrate, the Te-terminated 1x1 RHEED pattern remains unchanged. It implies that under the condition of $T_{Si} \geq T_{Te}$, the Te$_2$ molecular beam does not stick to the 1x1-Te surface and the film does not grow without simultaneous supply of the Bi atomic beam. This is important since it sets the lower limit of growth temperature for possible stoichiometric deposition. Below this limit, the Te atoms may no longer desorb from the 1x1-Te surface, leading to a situation similar to that in methods such as codeposition [23] or bulk crystal growth [5, 6]. The unique Te-rich growth dynamics and the criterion of $T_{Bi} > T_{Si} \geq T_{Te}$ observed here are closely analogous to the well-established GaAs MBE growth [25], confirming our initial hypothesis. Realizing that the only equilibrium phase of BiTe alloys is $Bi_2Te_3$ when $\theta \geq 3/2$ [26], which is always satisfied here ($\theta=8-20$), one easily understands why the growth of $Bi_2Te_3$ is strictly stoichiometric. This feature represents a fundamental difference between our method and those used in Refs. 5-6 and Ref. 23.

To illustrate the quality of the films, we carried out ARPES measurements. Figure 2 shows the band structure of the 80-nm-thick $Bi_2Te_3$ films taken along the $\overline{\Gamma}-\overline{M}$ direction. The film was grown with a $Te_2$/Bi flux ratio of 13 at a substrate temperature of 543 K. Besides the broad "M"-shape valence band (VB) at the bottom, a linear band dispersion representing the massless Dirac-like surface states (SS) is clearly seen in Fig. 2a. This feature is in agreement with theoretical predictions [4] and recent ARPES measurements on the cleaved (111) surface of the bulk crystal [5]. Most importantly, the bulk conduction bands (CBs) observed in the undoped $Bi_2Te_3$ [5, 6] crystals are completely absent at $E_F$, indicating that the as-grown films without doping are indeed an intrinsic topological insulator. Furthermore, the $Bi_2Te_3$ films are very stable in the UHV chamber, with no observable change in the ARPES spectra even after 72 hours. In comparison, for the bulk crystal doped with 0.67% Sn, the Fermi level shows an aging effect on the time scale of hours [5].

When the samples were continuously exposed to the photon beam for two hours, the Fermi level is observed to rise, as shown in Fig. 2c. The energy lift as estimated from the VB top is 50 meV. As a result, the bottom of the bulk CB is now located at 30 meV below $E_F$. The effect is shown more clearly in the momentum distribution curves (Fig. 2b and 2d). The lift enables us to estimate the energy gap, which is 170 meV (Fig. 2d) and consistent with both theory [4] and another ARPES experiment (165 meV) [4, 5]. Interestingly, if the sample is electrically grounded or warmed up to room temperature, the Fermi level falls back and the band structure shown in Fig. 2a is recovered. We conjecture that the lift is most likely due to optical doping induced charge accumulation [27]. The Fermi velocity along the $\overline{\Gamma}-\overline{M}$ direction from linear fitting (the red lines in Fig. 2a) is $3.32\times10^5$ m/s, which agrees well with the measurement on bulk materials ($3.87 \times 10^5$ m/s) [5], as well as the first-principles calculation ($3.5 \times 10^5$ m/s) [4]. Transport measurements on these thin films confirm the absence of bulk conduction. The resistivity of a 150 nm thin film shows metallic behavior from room temperature down to 1.5 K, presumably due to conduction from the topological surface states. The low temperature resistivity value, $\rho \sim 0.81$ mΩ·cm, is comparable to 0.65 mΩ·cm for the 0.67% Sn doped $Bi_2Te_3$ crystal that has been proved to be a topological insulator without bulk conduction [5].

Next, we study the thickness dependence of the topological features. In Fig. 3, we show the evolution of band structure of the $Bi_2Te_3$ films with a thickness from 1 QL to 5 QL. For the 1 QL film (Fig. 3a), there is a nearly parabolic free-electron-like band crossing the Fermi level and a gap (~0.5 eV) between CB and VB. The topological features start to appear in the 2 QL film, as shown in Fig. 3b. Except for an intense and broad bulk electron pocket on the top, the overall band structure is very similar to that of the intrinsic films shown in Fig. 2a. To understand the thickness-dependent electronic structure, we carried out first-principles calculations and found excellent agreement between our experiment and theory (lower panels of Fig. 3). The calculated surface state decay length is about 1 nm. Therefore, for the 1 QL film, the coupling between top and bottom surfaces is strong enough to open up a whole insulating gap. Starting from the 2 QL film, the inter-surface coupling becomes progressively weaker, and the topological features are recovered.

With increasing thickness, the Fermi level moves toward the bulk gap. The bulk CB is considerably smaller for the 5 QL film. This trend is more clearly revealed by the differential spectra maps (the middle panels of Fig. 3). When two surface layers couple to each other in an ultra-thin film, a hybridization gap is generally opened up. In this case, there is no clear way to define the concept of the 3D topological insulator. However, the hybridization gap of the 2D film can itself be topologically non-trivial, leading to topologically protected edge states, similar to the case of the HgTe quantum well [12, 13]. The layer-by-layer MBE growth of $Bi_2Te_3$ film enables us to investigate this question with fabricated devices.

We note that the major difference between experiment and theory is the position of $E_F$. Initially, we attributed the electronic pockets to ill-defined deposition conditions at the early stage of growth. However, the systematic upward shift of CB from Fig. 3b to Fig. 3e suggests that it is more likely due to Si segregation into the film. In Fig. 4, we summarize the Fermi velocity and energy gap of the five films. While the Fermi velocity exhibits a thickness-dependent behavior, the energy gap is less sensitive to thickness. The gap of the 3 QL film is already close to the bulk value (0.170 meV). Si doping may explain the evolution of $E_F$. If this is the case, the diffusion length of Si atoms at a given growth temperature will determine the thickness limit for obtaining an intrinsic topological insulator at that temperature.

In summary, we have established the Te-rich growth dynamics and criterion for preparing high quality topological films of $Bi_2Te_3$ by standard MBE technique. The information can apply to MBE growth of other topological materials such as $Bi_2Se_3$ and $Sb_2Te_3$ [4] and on other substrates. Our MBE-grown $Bi_2Te_3$ films not only reveal the intrinsic topological surface states in the cleanest manner reported so far, but also pave the road for future device applications involving topological insulators. The unique topological and thermoelectric properties of MBE-grown $Bi_2Te_3$ make it a strong candidate as a core component in multifunctional heterostructures for quantum computation, spintronics, and local chip cooling.

**Acknowledgments:** We would like to thank S. Q. Shen, Y. Ran and L. He for helpful discussions. The work in Beijing is supported by NSFC and the National Basic Research Program from MOST of China. The work at Stanford is supported by the Department of Energy, Office of Basic Energy Sciences, Division of Materials Sciences and Engineering, under contract DE-AC02-76SF00515. Y.L. is supported by DOD ARO under W911NF-07-1-0182, DOE under DE-FG02-04ER46159, and NSFC under 10628408.

**Author Information:** Correspondence and requests for materials should be addressed to J. F. Jia (jjf@mail.tsinghua.edu.cn).

**Figure Captions:**

FIG. 1 (color online). MBE growth of the intrinsic topological insulator of $Bi_2Te_3$. (a) The schematic crystal structure of $Bi_2Te_3$. (b) The bulk and the projected (111) Brillouin zones of $Bi_2Te_3$. (c) The RHEED patterns of the Si(111)-7x7 surface and the surface of 4 QL films taken along [11$\bar{2}$] direction. The RHEED pattern shows that the film has a well-defined 1x1 symmetry. (d) STM image of the 80- nm-thick film grown using the same condition in (c). All the steps seen in the image are ~10.17Å in height, corresponding to 1 QL. The image scale is 0.5 μm x 0.5 μm. The insert is an atomically resolved STM image (2.5 nm x 2.5 nm). (e) RHEED intensity of the (0, 0) diffraction versus growth time under different flux ratio and temperature. The growth was initiated by opening the Bi shutter. Upper panel: θ=9 (red), 13 (blue), 19 (green), $T_{si}$=543 K; Middle panel: θ=20 (red), 13 (blue), 8 (green), $T_{si}$=543 K; Lower panel: θ=13, $T_{si}$=563 K (red), 553 K (blue), and 543 K (green).

FIG. 2 (color online). Dirac-like band dispersion near the Γ point in the bulk Brillouin zone of the intrinsic topological insulator $Bi_2Te_3$ film. The film has a thickness of 80 nm (Fig. 1d). (a) ARPES intensity map of the film taken right the growth of the film. (b) Momentum distribution curves corresponding to the intensity map in (a). (c) ARPES intensity map of the film after continuous exposure of the photons for two hours. (d) Momentum distribution curves corresponding to the intensity map in (c). The surface states, valence band and conduction band of the bulk states are denoted SS, VB and CB, respectively. In (b) and (d), $E_F$, top of VB and bottom of CB are indicated by the blue, pink curve and green curve, respectively. The momentum distribution curves were measured at an interval of 10 meV. The red lines are guides to the eye for SS in the momentum distribution curves.

FIG. 3 (color online). Band structure of ultrathin films of $Bi_2Te_3$. (a) 1 QL. (b) 2 QL.

(c) 3 QL. (d) 4 QL. (e) 5 QL. All the spectra were taken along the $\overline{\Gamma}-\overline{M}$ direction. Upper panels: ARPES intensity maps; Middle panels: differential ARPES intensity maps; Lower panels: band structures from first-principles calculations. If the Fermi level (blue dashed line) in the calculated band structure (lower panels) is shifted to higher energy (red dashed line), the major features seen in the middle panels are well reproduced.

FIG. 4 (color online). Fermi velocity and gap of ultrathin films of $Bi_2Te_3$. The Fermi velocity was measured by fitting the experimental dispersion with $E=\hbar V_F K$, while the energy gap was estimated from the distance between the conduction band minimum and the valence band maximum as seen in Fig. 2 and Fig. 3.

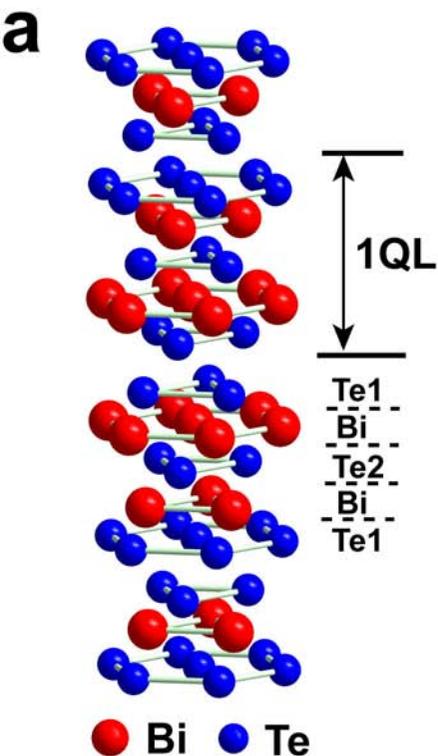
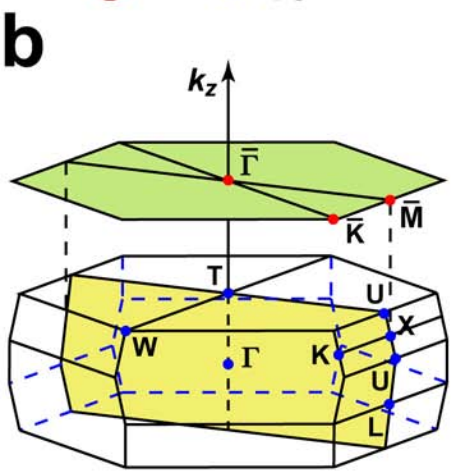
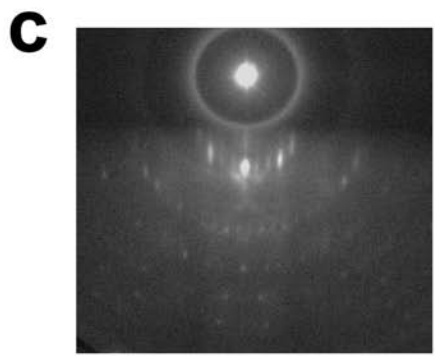
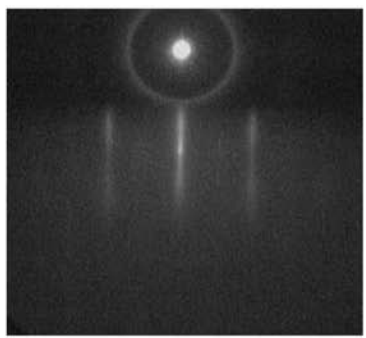
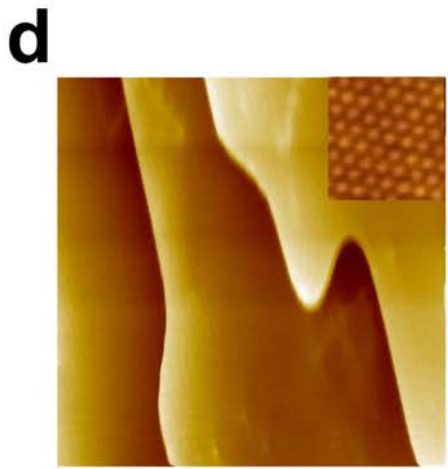
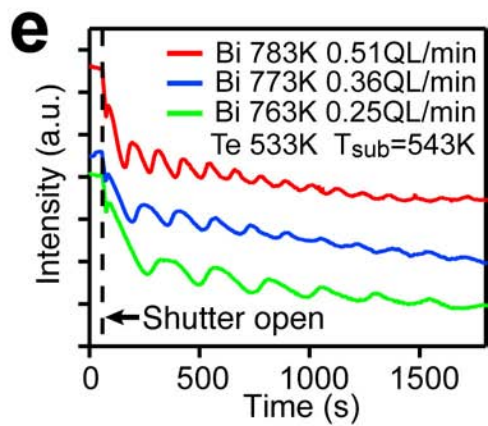
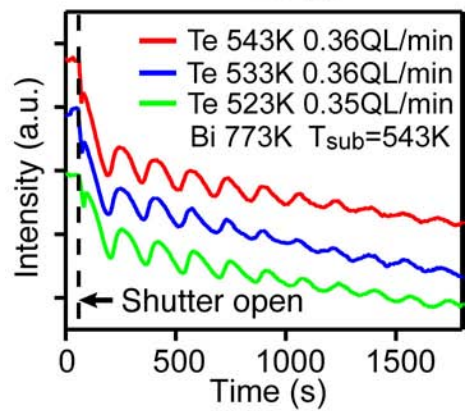
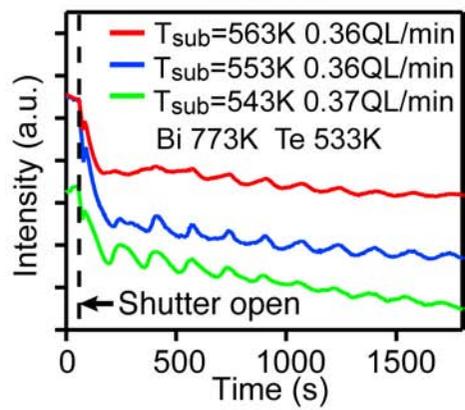

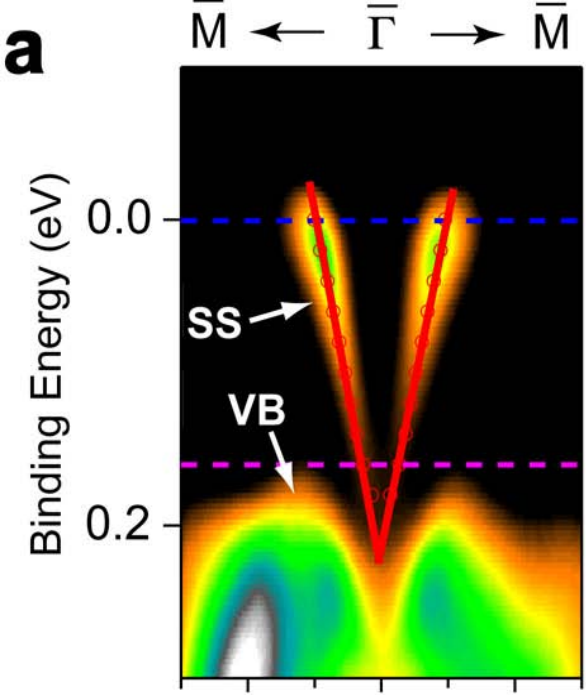
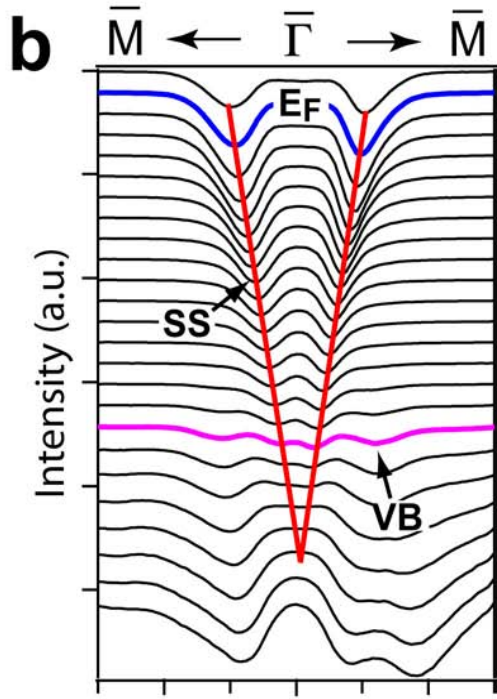
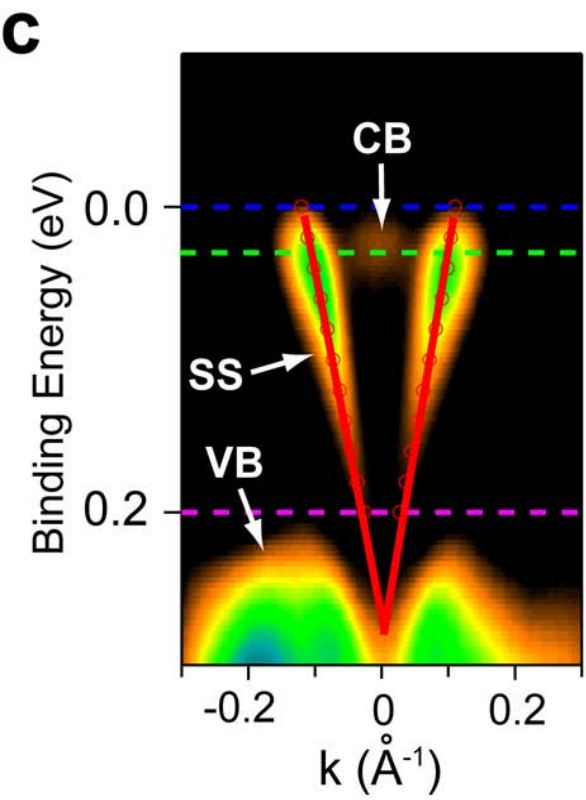
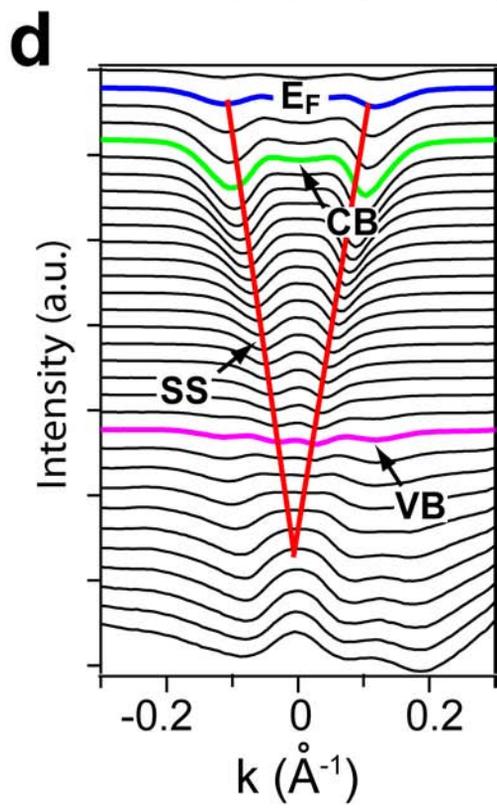

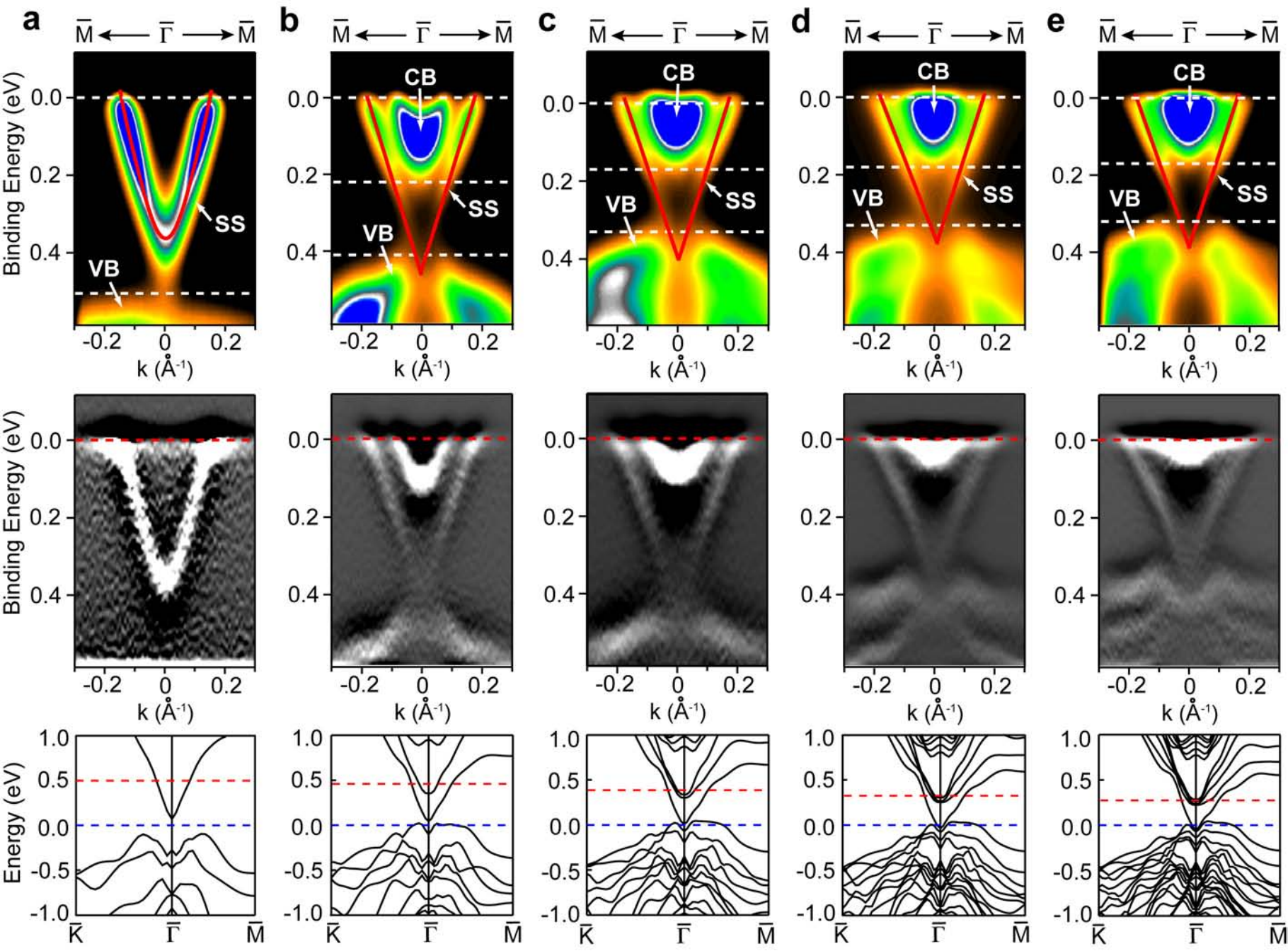

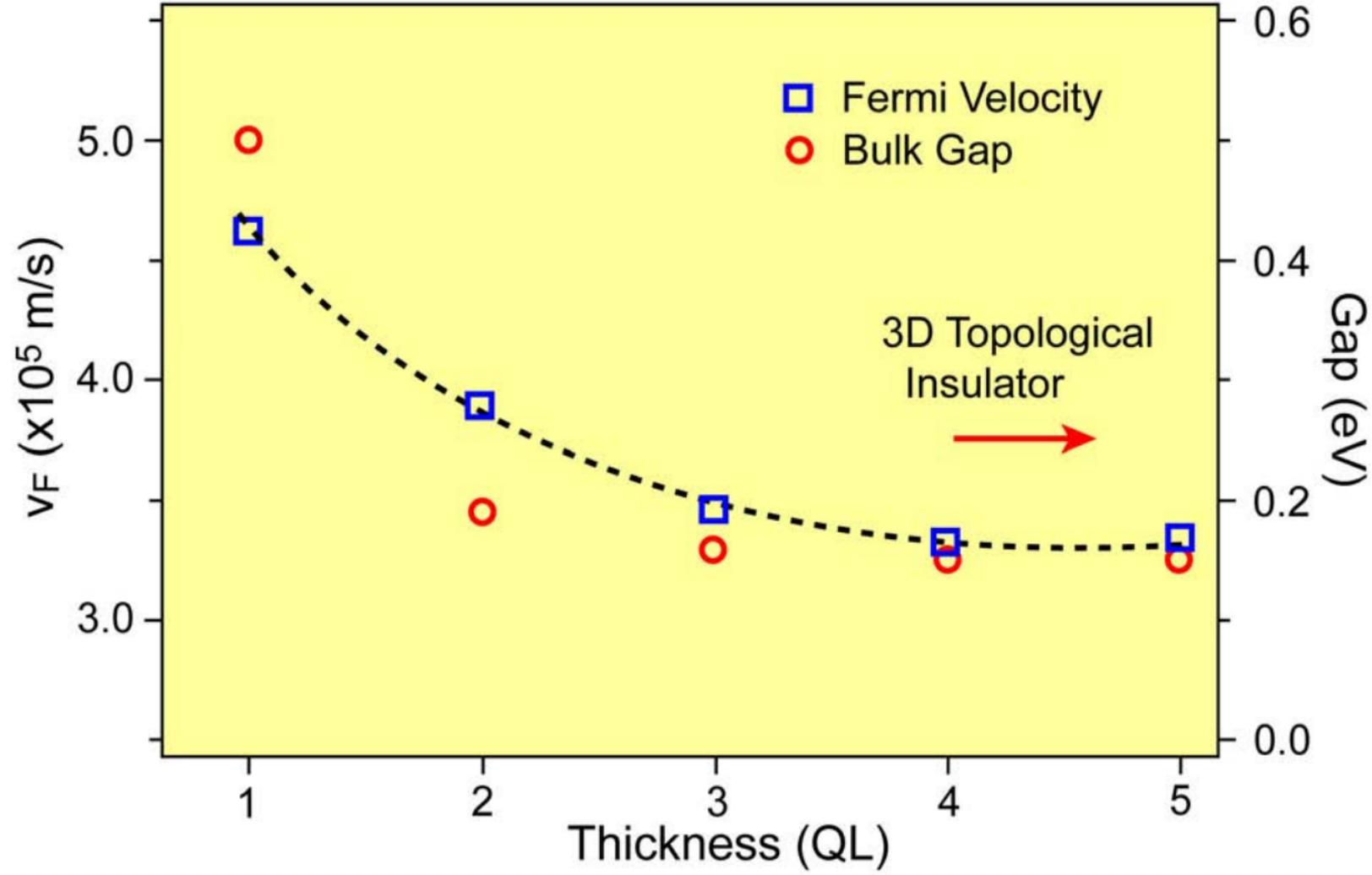